\documentclass{article}
\usepackage{graphicx}
\usepackage{subfigure}

\begin{document}
\title{Brane Configurations of BPS Domain Walls for the $\mathcal{N}=1^*$
$SU(N)$ Gauge Theory}
\author{Andrew Frey\\ {\it Department of Physics}\\ 
{\it University of California}\\ {\it Santa Barbara, CA 93106, USA}\\ 
{\it frey@physics.ucsb.edu}}
\maketitle
\abstract{We study supersymmetric domain walls in $\mathcal{N}=1$ $SU(N)$
gauge theory with 3 massive adjoint representation chiral multiplets.  
This theory, known as $\mathcal{N}=1^*$, can be obtained as a massive 
deformation of $\mathcal{N}=4$ Yang-Mills theory.  Following Polchinski and
Strassler, we consider the string dual of this theory in terms of spherical
5-branes and construct BPS domain walls interpolating between the many
vacua.  We compare our results to field theoretic domain walls and also 
find that this work is related to the physics of expanded ``dielectric''
branes near zero radius.}
\section{Introduction}\label{s:intro}
Recently, Polchinski and Strassler \cite{Polchinski:2000uf} 
found the string theory dual to an $\mathcal{N}=1$ gauge theory with 
adjoint matter, which can be obtained by giving masses to
three chiral superfields in the $\mathcal{N}=4$ $SU(N)$ gauge theory.  
They called this field theory $\mathcal{N}=1^*$ and obtained the string
dual by generalizing the AdS/CFT correspondence \cite{Maldacena:1998re}.  
On the string side of the duality, the $N$ D3-branes
that source the AdS geometry arrange themselves into 5-branes with $N$
units of D3 charge due to the RR 6-form corresponding to 
the mass perturbation in the CFT (as first discussed in \cite{Myers:1999ps}).  
Following Myers \cite{Myers:1999ps},
they suggested that the D3-branes form a 5-brane extended 
in the dimensions of the D3-branes with the other two dimensions wrapped on
an $S^2$.  The numbers and types of 5-branes in the configuration correspond
to specific vacua in the gauge theory, and their vacuum (coordinate) radii
and orientations are determined by a superpotential on the brane.  In 
particular, the totally Higgsed vacuum corresponds to a single D5-brane,
various Coulomb vacua correspond to multiple D5-branes, and the confining and
oblique confining vacua correspond respectively to a single NS5- or 
$(1,k)$5-brane (for $1\leq k <N$).
For specifics of the brane description, see \cite{Polchinski:2000uf}; other
studies of the $\mathcal{N}=1^*$ theory include
\cite{Vafa:1994tf,Donagi:1996cf,Strassler:1997ny,
Dorey:1999sj,Dorey:2000fc,Aharony:2000nt}.

Because the $\mathcal{N}=1^*$ $SU(N)$ theory has a large (but finite for
finite $N$) number of vacua, there should generically be domain walls 
interpolating between pairs of those vacua.  In supersymmetric theories with
domain walls, there has been much interest in finding BPS domain wall
solutions -- domain walls that preserve some supersymmetry.  In particular,
for gauge theories, studies of BPS domain walls include 
\cite{Kovner:1997ca,Kogan:1998dt,Shifman:1998vf,Dvali:1999pk,Gabadadze:1999pp,
Witten:1997ep,Smilga:1997pf,Smilga:1997cx,Smilga:1998vs,deCarlos:1999xk}; see 
\cite{Shifman:1998vf} for further references on BPS domain walls.  Recently,
\cite{Dorey:2000fc,Bachas:2000dx} studied BPS domain walls in the 
$\mathcal{N}=1^*$ theory from the field theory perspective.  

In this paper, we investigate the brane configurations corresponding to 
BPS domain walls that interpolate between different vacua in the 
gauge theory.  On the string side, the 5-branes bend from one vacuum state to
the other, and when branes with non-zero net 5-brane charge intersect, another
5-brane fills the $S^2$ at the intersection \cite{Polchinski:2000uf}.  
Polchinski and Strassler \cite{Polchinski:2000uf} discussed two examples of
domain walls in the small coupling limit and compared the vacuum state
superpotentials and domain wall tensions to
exact field theoretic calculations of \cite{Dorey:1999sj,Dorey:2000fc},
finding agreement within their approximations.  

We expand those results to finite string coupling and construct
domain wall brane configurations.  In section \ref{s:actions}, we review
the 5-brane actions of \cite{Polchinski:2000uf} and establish some 
approximations.  In section \ref{s:forces}, we find conditions necessary
for the mechanical equilibrium and supersymmetry of the 5-brane junctions,
and section \ref{s:general} uses the general results of sections 
\ref{s:actions} and \ref{s:forces} to confirm that the BPS bound for the 
domain wall tension on the string side matches the field theoretic bound.
In section \ref{s:specific}, we construct a number of BPS domain walls and
note interesting examples.  Finally, in section \ref{s:summary}, we 
summarize our results and discuss their relation to the body of research on
BPS domain walls in field theory.  In both sections \ref{s:specific} and
\ref{s:summary}, we will discuss the supersymmetric minimum in the brane
potential at vanishing 5-brane sphere size \cite{Polchinski:2000uf,
McGreevy:2000cw}
and its relation to the brane picture of BPS domain walls.  In the end, though,
we will have to confess ignorance as to the meaning or existence of a zero
size state for the 5-brane spheres.

While preparing this paper, we became aware of work by C. Bachas, J. Hoppe,
and B. Pioline \cite{Bachas:2000dx} that has some overlap with this work.
Specifically, they find BPS domain wall configurations interpolating between
Coulombic vacua (and the Higgs vacuum) within field theory.  We discuss these
domain walls in section \ref{s:onecharge} and compare our results to those of
\cite{Bachas:2000dx} in section \ref{s:compare}.

\section{Brane Actions}\label{s:actions}
In this section, we will discuss the action that describes the 5-brane 
bending for the domain walls.  Through the rest of this paper, we will follow
the conventions of Polchinski and Strassler \cite{Polchinski:2000uf},
working to leading order in their small parameter, the ratio of 5-brane to 
D3 charge.  In doing so, we ignore the near-shell corrections to the metric
and supergravity fields, 
so we take the dilaton to be constant and the Einstein frame
metric to be equal to the string frame metric.

First, we rederive the action for brane bending given in equation (126) 
of \cite{Polchinski:2000uf}.
The $S^2$ part of the 5-brane twists and
contracts (or expands) as we pass from one vacuum state to another (moving
in, without loss of generality, the $x^1$ direction with translational 
invariance in the $x^2$ and $x^3$ directions).  To start, we note that the
induced metric in the directions parallel to the D3-branes is
\begin{equation}
G_{\mu\nu}(\mbox{induced}) = G_{\mu\nu} + G_{mn}\partial_\mu x^m \partial_\nu
x^n\, .
\end{equation}
The Dirac-Born-Infeld action of a $(c,d)$ 5-brane is therefore
\begin{eqnarray}
S &=& -\frac{\mu_5}{g |M|^2} \int d^6 \xi \left[ -\det \left( |M| 
Z^{-1/2} \eta_{\mu\nu} + |M| Z^{1/2}\delta_{mn} 
\partial_\mu x^m \partial_\nu x^n\right) \right. \nonumber \\
& & \left. \times \det\left( |M| G_\perp + 2\pi \alpha^\prime \mathcal{F} 
\right) \right]^{1/2}\, .
\end{eqnarray}
Here, $M = c\tau + d$ with coupling $\tau = \theta/2\pi + i/g$ for the $(c,d)$
brane, and $G_\perp$ is the metric on the wrapped $S^2$ of the 5-brane
\cite{Polchinski:2000uf}.

Under the assumption of slow bending (small derivatives of brane position),
the factor in the $x^{0-3}$ directions gives
\begin{equation}
|M|^2 \left( Z^{-1} + \frac{1}{2} \eta^{\mu\nu} \partial_\mu x^m 
\partial_\nu x^m \right) = |M|^2 \left( Z^{-1} + (2\pi \alpha^\prime)^2
\eta^{\mu\nu} \partial_\mu \bar{\phi} \partial_\nu \phi \right)\, .
\end{equation}
In the last step, we have substituted the scalar field $\phi$ introduced in 
\cite{Polchinski:2000uf} to describe the size and orientation of the $S^2$.  
For a real unit vector $e^i$ on the $S^2$, $\phi$ is defined by
\begin{equation}
(2 \pi \alpha^\prime) \phi e^{1,2,3} = \frac{1}{\sqrt{2}} \left( x^{4,5,6} +
i x^{7,8,9} \right)\, .
\end{equation}
This may seem like a somewhat suspect expansion, given that $Z$ diverges at
the brane.  However, it corresponds to the self-reaction of a charge, which 
should be ignored, as discussed in \cite{Polchinski:2000uf}.  
We can think of the 5-brane as built
up from infinitesimal 5-branes, each of which acts as a probe brane to the
rest of the geometry.  Since 
each probe action is independent of $Z$, the expansion follows for the full
5-brane.

The rest of the action follows as in \cite{Polchinski:2000uf}; 
we can expand both the $S^2$ 
determinant and the Chern-Simons action in powers of the D3 charge, which is
assumed to dominate.  Then the action for $n$ D3 charges (integrated over the
$S^2$) is
\begin{equation}\label{action}
S = -\int d^4 x \left[ \frac{n \mu_3}{g} (Z^{-1}-Z^{-1}) + \frac{n}{2\pi g}
\eta^{\mu\nu} \partial_\mu \bar{\phi} \partial_\nu \phi + \frac{2\pi g}{n}
|W_\phi |^2 \right]
\end{equation}
where $W_\phi = 1/(2\pi g) (mn \phi +i 2\sqrt{2} M \phi^2 )$ is the derivative
of the superpotential
\begin{equation}\label{superpotential}
W = \frac{1}{2 \pi g} \left( \frac{mn}{2} \Phi^2 + i \frac{2\sqrt{2}}{3} M 
\Phi^3 \right)\, .
\end{equation}
A few comments are in order.  First, the leading $Z^{-1}$ terms from the 
Dirac-Born-Infeld and Chern-Simons actions cancel, as they should for parallel
D3-branes.  However, when we consider the tension of the 5-brane, it is
precisely the D3 tension $n\mu_3 / g$ that dominates.  Also,
the vacuum configuration of the 5-branes occur at the roots of the 
potential $|W_\phi|^2$, where the configuration and superpotential are
\begin{equation}\label{vacua}
\phi_v = \frac{imn}{2\sqrt{2} M}\, ,\, W_v = -\frac{m^3 n^3}{96 \pi g M^2}\, .
\end{equation}
For a configuration with multiple branes, the superpotential is summed over the
branes, and the vacuum configuration of each brane is unaffected by the other
branes \cite{Polchinski:2000uf}.  

Since we hope to find supersymmetric domain walls (with configurations
varying only in the $x^1$ direction), it is useful to write
the action (eqn. \ref{action}) as
\begin{equation}\label{bpsaction}
S = -\int dx \left( 
\frac{2\pi g}{n} \left| \frac{n}{2\pi g} \partial_1 \phi - \Omega
\overline{W_\phi}\right|^2 +\Omega \partial_1 \overline{W} +\overline{\Omega}
\partial_1 W\right)\, ,
\end{equation}
with $\Omega$ a complex phase.  From the above, we can see that brane bending
that follows the ``BPS equation''
\begin{equation}\label{bps}
\frac{n}{2\pi g} \partial_1 \phi = \Omega \overline{W_\phi}
\end{equation}
both satisfies the equations of motion and preserves supersymmetry.  In this
case, the domain wall tension due to the brane bending is given by the surface
terms.  Further, each 5-brane in the domain wall must bend with the same
phase $\Omega$ for supersymmetry to be preserved.  Also, since the BPS
equations for different 5-branes (in the same vacuum state, say) are decoupled
except for having the same value of $\Omega$,
each 5-brane bends independently.   We emphasize that each 5-brane follows
its own BPS bending equation independently of any other 5-branes present
and note that we will consider only domain walls with BPS brane bending.

At some point in the transition from one vacuum to another with a differently
charged 5-brane(s), charge conservation requires that 
an additional 5-brane be present, as discussed in \cite{Polchinski:2000uf}.  
If the domain wall
interpolates between, for example, a $(c,d)$5-brane and a $(c^\prime,
d^\prime)$5-brane, then the third 5-brane has charge $(c^\prime-c,d^\prime-d)$.
Further (taking the vacua to change along the $x^1$ direction), this
extra 5-brane extends in the $x^{2,3}$ directions and fills the $S^2$ at the
intersection of the two ``vacuum branes.''

The DBI part of the action of this 5-brane ball is (at lowest order in
the perturbations)
\begin{equation}
S = -\frac{\mu_5}{g|M|^2} \int d^6\xi \sqrt{-\det \left( |M| G_{ab}\right)} 
= -\frac{\mu_5 |M|}{g} \int d^3x \frac{4\pi r_0^3}{3}\, ,
\end{equation}
where $r_0$ is the radius of the $S^2$
and $M=c\tau+d$ is the 5-brane charge as above.  
The metric factors $Z$ have canceled
because the 5-brane is extended in 3 dimensions each of factor $Z^{-1/2}$ and 
$Z^{1/2}$.  Then the domain wall tension due to this ball-filling brane is
\begin{equation}\label{Sfill}
\tau(\mbox{ball})=\frac{\mu_5 |M|}{g} \frac{4\pi r_0^3}{3}=
\frac{2\sqrt{2}|\phi_0|^3 |M|}{3\pi g}
\end{equation}
for $\phi_0$ the configuration for all the vacuum branes at the brane junction.

\section{Force-Balancing Equations}\label{s:forces}
We consider in this section the force-balancing equations at the 5-brane
junction, which are needed to maintain 
mechanical equilibrium and preserve supersymmetry.  These are 
analogous to conditions derived in \cite{Aharony:1997ju,Aharony:1998bh}
but are quantitatively
different due to the D3 tension here.  We specialize to a 
static domain wall along $x^1$ (hereafter $x$) and translationally invariant
in $x^{2,3}$.  We consider any number of
5-branes labeled by the index $i$ with 5-brane charges $M_i$ and $n_i$ units
of D3 charge each approaching the junction from smaller $x$ and 5-branes
labeled by $j$ approaching from larger $x$.

For a 5-brane junction to occur, the 5-branes must first intersect.  On first
glance, this means that the brane configuration variables $\phi$ must be
the same for all the vacuum 5-branes at the junction.  However, we note that
$\phi$ and $-\phi$ describe the same sphere, so there could be a brane junction
between, for example, two 5-branes with opposite $\phi$.  We deal with this 
case by noting that the action (\ref{bpsaction}) (and BPS equation (\ref{bps}))
is invariant under $\phi \rightarrow -\phi,\, M \rightarrow -M$, so a 5-brane
is equivalent to the corresponding anti-5-brane with opposite configuration
variable.  Physically, taking $\phi\rightarrow -\phi$ leaves the $S^2$
invariant while reversing its orientation and therefore the 5-brane charge.
Then we can always choose 5-brane charges at a junction so that all
the 5-branes have the same configuration variable, which takes the
value $\phi_0$ at the junction (though we might need to consider different
junctions separately in each domain wall).  

The force-balancing conditions should be derived in
an inertial reference frame.  In
this flat metric, the D3- and 5-brane tensions are simply the flat space
values and give the force per proper area on the junction in the directions 
of extent of those branes.  Thus, for $n$ D3-branes on an $S^2$ of 
coordinate radius $r$, the total D3 tension on a unit area of the $S^2$ is
\begin{equation}
\frac{\mu_3}{g} \frac{n}{4\pi r^2 Z^{1/2}} \equiv \frac{\mu_3}{g} \rho\, .
\end{equation}
Technically, we are working slightly away from the 5-branes (so $Z$ is finite)
-- this substitutes for building up the domain wall out of infinitesimal
charges -- in an orthonormal basis aligned along the coordinate axes.  We 
consider the north pole of the spherical brane junction, which is in the
$x^6, x^9$ plane with $\arg (x^6+i x^9)=\arg (\phi_0) \equiv \alpha$.  
Henceforth, we denote $x^{6+i9} \equiv x^6+i x^9$.

Now consider a unit vector $v_{i,j}$ along each vacuum brane in the direction
out of the junction.  In the small bending approximation,
\begin{equation}
v_{i,j}^{\mathbf{x}} = \mp 1\ ,\ v_{i,j}^{\mathbf{6+i9}} = \mp Z^{1/2}
\partial_x x_{i,j}^{6+i9}
\end{equation}
where bold indices are the orthonormal frame indices and the signs are for
the indices $i,j$ respectively.  A ball-filling 5-brane has a tangent
vector 
\begin{equation} 
v^{\mathbf{6+i9}} = -e^{i\alpha}
\end{equation}
leaving the north pole of the brane junction.  Then mechanical equilibrium
requires
\begin{equation}\label{forces}
\frac{\mu_5}{g} \left| \Delta M\right| v +\frac{\mu_3}{g}\left( \sum_i 
\rho_i v_i +\sum_j \rho_j v_j \right)=0
\end{equation}
for $\Delta M = \sum_j M_j - \sum_i M_i$.  Equation (\ref{forces}) gives
the conditions
\begin{equation}\label{d3cons}
\sum_i n_i = \sum_j n_j
\end{equation}
in the $\mathbf{x}$ direction and
\begin{equation}\label{angle}
\left| \Delta M \right| e^{i\alpha} = \frac{1}{4\pi \sqrt{2} \left|\phi_0
\right|^2} \left(\sum_j n_j \partial_x\phi_j - \sum_i n_i \partial_x
\phi_i \right)
\end{equation}
in the $\mathbf{6+i9}$ directions.

The first of these equations simply gives conservation of D3 charge across
the junction.  For BPS brane bending according to equation (\ref{bps}),
D3 charge conservation requires the linear terms from the potential
to cancel.  Then we find simply that
\begin{equation}\label{cond}
e^{i 3\alpha} = -i \Omega \overline{\Delta M}
/\left|\Delta M\right|\, .
\end{equation}
This implies that the discontinuity of the superpotential at the junction is
\begin{equation}\label{phase}
\Delta W_0 = \frac{i}{2 \pi g} \frac{2\sqrt{2}}{3}  \Delta M \phi_0^3
= \Omega |\Delta W_0|\, .
\end{equation} 
Note, however, if there is no 5-brane ball at the junction, there is no
condition on the phase of $\phi_0$.

To this point, we have considered only positive D3-brane charge, as negative
D3 charges would break supersymmetry and add a high energy cost.  We can
also see this from mechanical equilibrium considerations; if we allow
negative D3 charges, equation (\ref{d3cons}) becomes
\begin{equation}
\sum_i \left| n_i\right| = \sum_j \left| n_j \right|
\end{equation}
which conflicts with charge conservation at the junction.  Thus, we 
conclude that negative D3 charges will not appear in any BPS domain wall.

\section{Domain Walls: Generalities}\label{s:general}
In this section, we discuss the BPS bound for the domain wall tension and
confirm that it matches the BPS bound from field theory for a domain
wall interpolating between any two vacua of the $\mathcal{N}=1^*$ theory.  We
also make some observations that will allow us to discuss some specific
explicit domain wall solutions in the following section.

As discussed above, we consider domain walls which follow the BPS equation 
(\ref{bps}) for bending of the 5-branes.  Combining the BPS equation and
its conjugate, we find
\begin{equation}
\overline{\Omega} \partial_x \phi\, W_\phi = \Omega \partial_x \overline{\phi}
\,\overline{W_\phi}\, ,
\end{equation}
or that the imaginary part of $\overline{\Omega} W$ is conserved.  
This implies that,
up to discontinuities in $W$ at brane junctions, 
the BPS trajectory follows a straight line in
the complex $W$ plane with a tangent vector of complex phase $\pm\Omega$
(in the direction of increasing $x$).  Due to the force-balancing condition 
equation (\ref{phase}), 
the discontinuous jump in $W$ at a brane junction also has the phase $\Omega$, 
so the trajectory of the superpotential, including discontinuities,
is then a straight line directly from $W(x=-\infty)$ to $W(x=\infty)$ (ie,
between the two vacuum values).  Thus we have $\Omega=\pm\Delta W/|\Delta W|$.
In the case that there is no ball-filling 5-brane at a junction, there
is no constraint from mechanical equilibrium, but there is also no 
discontinuity in the superpotential, so we still have a straight line.
We should note that, in field theory, 
the superpotential trajectory would 
also follow this straight line but without discontinuities.

Now consider the domain wall tension.   
From the action (\ref{bpsaction}) and BPS equation (\ref{bps}), 
the tension from brane bending is 
\begin{equation}
\int dx \left(\overline{\Omega} \partial_x W 
+ \Omega\partial_x \overline{W}\right)= 2\mbox{Re} \left(\overline{\Omega}
\Delta W\right)
\end{equation}
along any single 5 brane.   We see that $\Omega=-\Delta W/|\Delta W|$ gives
a negative tension, which comes from a {\em positive definite} Hamiltonian,
so that choice is unphysical.  Taking $\Omega = \Delta W/|\Delta W|$, 
the brane bending contributes
$2|\Delta W(\mbox{bending})|$ to the domain wall tension.  Further, twice the 
discontinuity in the superpotential at a 5-brane junction is equal in 
magnitude to the tension of the 5-brane ball at that junction:
\begin{equation}
2 |\Delta W_0| = \frac{4\sqrt{2}}{3(2\pi g)} |\Delta M| \left| \phi_0\right|^3
= \tau (\mbox{ball})
\end{equation}
(compare to eqn. (\ref{Sfill})).  Thus, assuming 
$\Delta W/|\Delta W|=+\Omega$,
we find the BPS bound for domain wall tension
\begin{equation}\label{tension}
\tau_{DW} = 2 \left| \Delta W \right|
\end{equation}
in agreement with field theoretic results since \cite{Polchinski:2000uf} 
found that the vacuum superpotentials (eqn. (\ref{vacua})) match the exact
field theoretic results (within our approximations).  
Since the BPS bound on the 
domain wall tension follows from the supersymmetry algebra
(see, for example, \cite{Shifman:1998vf} for a review of the central charges
of domain walls), it is not surprising to find the same bound; however,
it is satisfying to see new physics give the same tension.

\section{Domain Walls: Specifics}\label{s:specific}
In this section, we will discuss explicit domain wall solutions and comment
on them.  We will begin by discussing domain walls between vacua with only
one type of 5-brane (for example, D5-branes only on both the left and the
right), which we can discuss analytically because the 5-brane spheres
keep the same orientation throughout the domain wall.  
We will proceed to consider
domain walls between vacua with single 5-branes of different types and will
finally discuss some examples of domain walls between vacua with multiple
5-branes of different types.  Throughout, we take the mass $m$ to be positive.

\subsection{Single Type of 5-Brane Charge}\label{s:onecharge}
Here, we consider only domain walls between vacua with only one type of 
5-brane charge given by $M=c\tau +d$.  As discussed above, each 5-brane has
a vacuum configuration and superpotential given by equation (\ref{vacua})
(inserting the appropriate value of $n$ for each 5-brane).  Therefore, we see
that all the 5-branes have the same orientation and same phase of the 
superpotential.  For definiteness, we will always take domain walls with the
magnitude of the superpotential increasing from negative to positive $x$,
which gives
\begin{equation}\label{omega}
\Omega = -\exp\left(-2 i \arg M\right)\, .
\end{equation}
Additionally, we can rotate the phase of the brane configuration variables
to $\phi = i \exp(-i \arg M) \psi$ with vacuum  $\psi_v =
m n /2\sqrt{2} |M|$.  Then the BPS equation becomes
\begin{equation}\label{realbps}
\frac{1}{m} \partial_x \psi = \overline{\psi} - \frac{\overline{\psi}^2}
{\psi_v}\, .
\end{equation}
Without losing generality, we can take all the vacuum 5-branes to be D5-branes
from this point forward; the only difference would be the size of $\psi_v$.

\begin{figure}[t]
\centering
\includegraphics[scale=0.8]{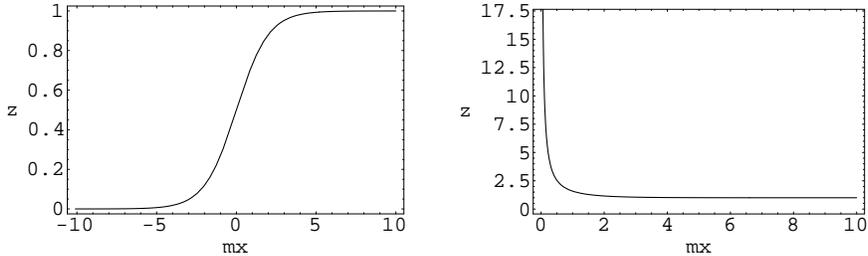}
\caption{\label{f:realsol}These are the two positive, real solutions to 
equation (\ref{realbps}) given by equation (\ref{lesssol}) on the left and
equation (\ref{greatsol}) on the right.  The vertical axis in both cases
is $z=\psi/\psi_v$.}
\end{figure}

To begin, we consider domain walls in which $\psi$ remains real for all the
5-branes, so the BPS equation has the following solutions:
\begin{eqnarray}
\psi_< (x,x_0,n) &=& \psi_v \frac{e^{m(x-x_0)}}{1+e^{m(x-x_0)}}\  \mbox{for}\,
0< \psi < \psi_v\, ,\,  \mbox{all}\, x \label{lesssol}\\
\psi_> (x,x_0,n) &=& \psi_v \frac{e^{m(x-x_0)}}{e^{m(x-x_0)}-1}\ \mbox{for}\,
 \psi > \psi_v\, ,\,  x>x_0 \label{greatsol}\\
\psi_- (x,x_0,n) &=& \psi_v \frac{e^{m(x-x_0)}}{e^{m(x-x_0)}-1}\ \mbox{for}\,
 \psi < 0 \, ,\,  x<x_0\, .
\end{eqnarray}
In the above, $x_0$ is an integration constant, and the vacuum value $\psi_v$
is the appropriate value for a 5-brane with $n$ D3 charges.  The two 
positive solutions 
are plotted in figure \ref{f:realsol}.  
It is important to note that none of these solutions
goes to $\psi_v$ as $x\rightarrow -\infty$, so all the vacuum D5-branes at
negative $x$ must remain in their vacuum configurations until they reach a
brane junction (for the real solutions that we consider here).  We will come
back to this point later.  Also, the force-balancing condition (eqn.
(\ref{cond})) for $\psi >0$ implies that any 5-brane junction should have 
more D5-branes on the left (lesser $x$) than on the right (or an equal number) 
in a BPS configuration.

Another useful solution to the BPS equation is that
for a sphere of D3-branes with no 5-brane charge (henceforth a 
``zero-5-brane''), which bends according to the BPS equation with $M = 0$.
With $\Omega$ and $\psi$ defined as above, the BPS equation for a zero-5-brane
and its real solution are
\begin{eqnarray}
\partial_x \psi &=& m \overline{\psi}\\
\psi_{(0)}(x,\psi(0)) &=& \psi(0) e^{mx}\, .
\end{eqnarray}
A few words are necessary about the zero-5-brane configuration.  In a vacuum
state, the D3 branes with no 5-brane charge would collapse to a point.  This 
situation seems physically similar 
to the minimum of the potential $|W_\phi|^2$ at $\phi=0$ for
any 5-brane charge, which is usually considered an unphysical minimum 
\cite{Polchinski:2000uf,McGreevy:2000cw}.  However, inside a  domain wall, 
we need not be  concerned whether the
zero-5-brane corresponds to a physical vacuum or not.  We will discuss this
point further below.

Now we can construct explicit domain wall solutions.  Because we are taking
all the vacuum 5-branes to be D5s, we are discussing
domain walls between various Coulomb vacua and the Higgsed vacuum -- 
Coulomb vacua are vacua with multiple D5-branes and $U(1)^k$ or $SU(k)$
gauge symmetry unbroken, while the Higgs vacuum is a single D5-brane with
all gauge symmetry broken.  
In the following domain walls, we will define $\psi_V = mN/2\sqrt{2}|M|$,
the vacuum configuration for a single 5-brane vacuum, for convenience.

\begin{figure}[t]
\centering
\includegraphics[scale=0.7]{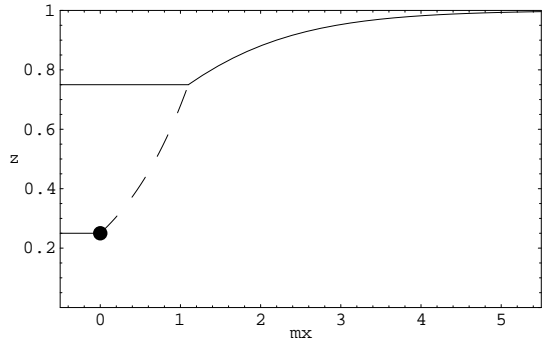}
\caption{\label{f:twoone}The domain wall between two D5s and one D5.  In this
case, $f=3/4$, and in all figures, $m=1$ and $N=12000$.  
The vertical axis is 
$z=\psi/\psi_V$.  The dashed line is a zero-5-brane, and the dot is a 
ball-filling D5-brane. (These conventions are standard in figures 
\ref{f:twotwo} and \ref{f:threetwo}.)}
\end{figure}

The first
domain wall we will consider is between a vacuum with two D5 branes with
$fN$ and $(1-f)N$ D3 charges each ($1/2\leq f< 1$) and a single D5 with all
$N$ D3 charges.  In this domain wall, the smaller D5 on the left remains
in its vacuum state for $x<0$, where there is a junction with a ball-filling
D5-brane and a zero-5-brane that follows the solution 
$\psi_{(0)}(x,(1-f)\psi_V)$.  The larger vacuum brane on the left stays in
its vacuum state for $x<(1/m) \ln(f/1-f)$, where it enters a junction with 
the other end of the zero-5-brane and the vacuum brane from the right.  The
single D5-brane on the right follows $\psi_<(x,0,N)$ for $x>(1/m) \ln(f/1-f)$.
This configuration is shown in figure \ref{f:twoone}.  
We should note that when $f=1/2$, there is no zero-5-brane.

\begin{figure}[t]
\centering
\subfigure[BPS Domain Wall]{\label{f:twotwo:a}
\includegraphics[scale=0.58]{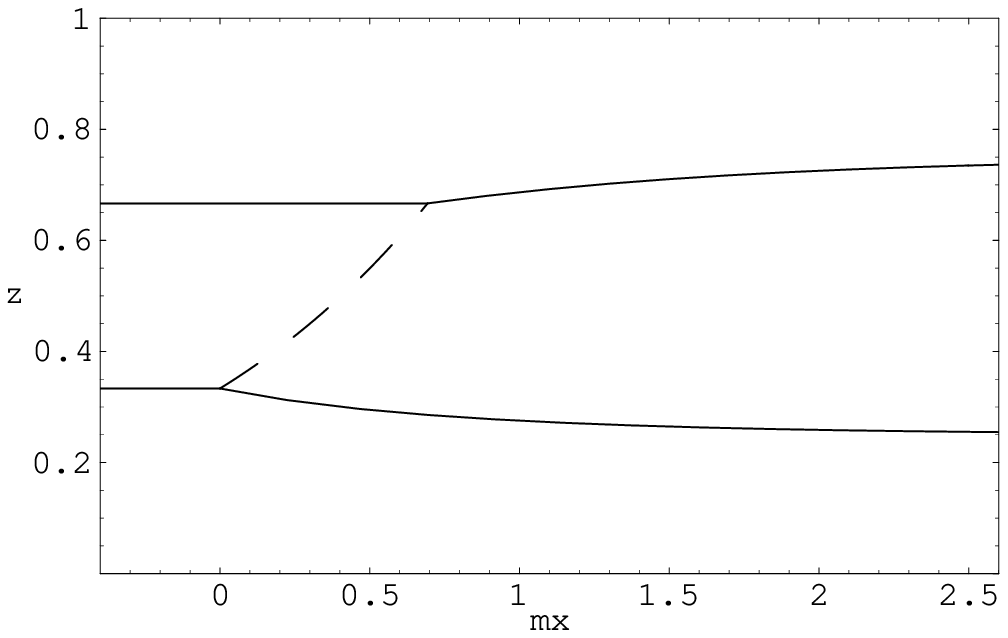}}
\subfigure[Unstable, Non-BPS Domain Wall]{\label{f:twotwo:b}
\includegraphics[scale=0.58]{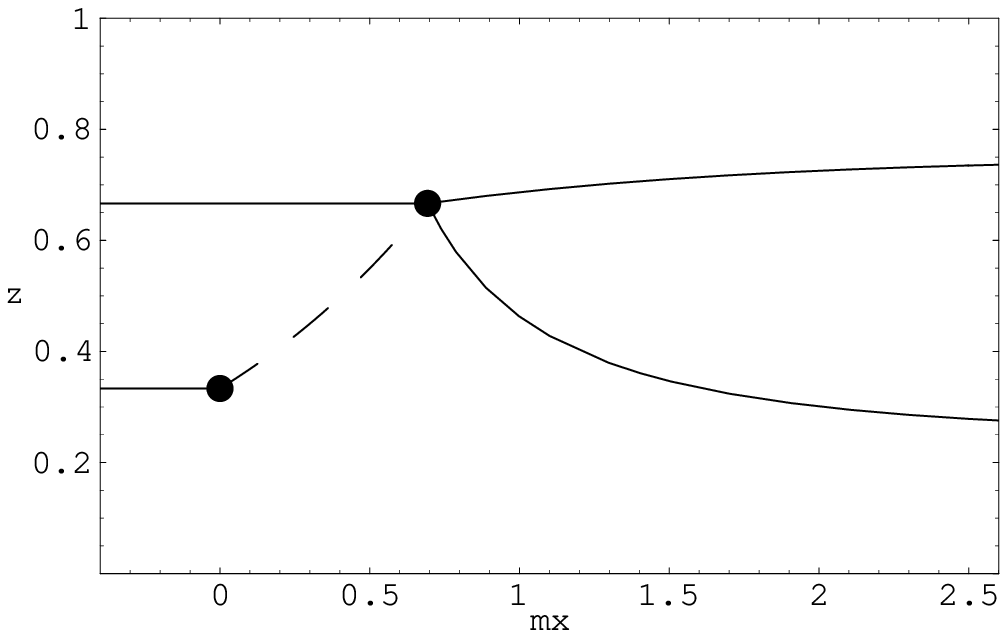}}
\caption{\label{f:twotwo}Two domain wall configurations between two
Coulomb vacua.  In this case, $f_1=2/3$ and $f_2=3/4$.  There is also
a continuous family of non-BPS domain walls interpolating between these
two cases, which allow the non-BPS domain wall to decay into the BPS domain
wall.}
\end{figure}

Another domain wall of interest is that between two Coulomb vacua, such
as the vacuum discussed above, with D3 charges given by $f_1$ on the left
and $f_2$ on the right.  Because the magnitude of the superpotential of such
a vacuum is proportional to $f^3 + (1-f)^3 = 1-3f+3f^2$, we see that we should
take $f_2>f_1$ to have the larger superpotential at positive $x$.  The BPS 
domain wall corresponding to these vacua is shown in figure
\ref{f:twotwo:a}.  It has no
5-brane balls and has brane junctions between the two smaller D5s and between
the two larger D5s, which are connected by a zero-5-brane.  If we choose to
have the junction of the two smaller branes at $x=0$, the small D5 on the right
bends according to $\psi_> (x,x_0,(1-f_2)\psi_V)$ for $x>0$, the zero-5-brane
bends according to $\psi_{(0)}(x,(1-f_1)\psi_V)$ for $0<x<y=(1/m)\ln 
(f_1/(1-f_1))$, and the large D5 on the right bends according to
$\psi_< (x,x_0,f_2\psi_V)$ for $x>y$.  Here, $x_0 = -(1/m)
\ln ((1-f_1)/(f_2-f_1))$.

Some interesting physics arises if we consider other domain walls 
interpolating between these two vacua.  If we just consider BPS brane
bending (as in eqns. (\ref{lesssol},\ref{greatsol})), 
it appears that we could have BPS domain walls such
as those shown in figure \ref{f:twotwo:b}, 
where the small D5 on the right does not connect
to the small D5 on the left.  In fact, it appears that there is a continuous
family of such domain walls with a D5-brane branching off of a zero-5-brane.  
However, these domain walls are {\em not} BPS and are not mechanically
stable (despite the fact that they have BPS brane bending) 
because the brane junction involving the small vacuum brane on the right
{\em does not satisfy the force-balancing condition} (\ref{cond}).
These domain walls actually have a higher tension than the BPS bound
due to ``backtracking'' of the domain wall trajectory in the complex $W$ plane.
Essentially, the smaller D5-brane at positive $x$ has to bend too much.
So, given one of these non-BPS domain walls, the D5-brane balls, which are
anti-branes of each other, are free to attract (because there is a continuous
family of domain walls) and annihilate, leaving the BPS solution, classically
at least.

\begin{figure}[t]
\centering
\subfigure[The three D5-branes carry 1/6, 1/3, and 1/2 of the D3 charges;
the two carry 1/4 and 3/4.  Note that the two zero-5-branes carry different
amounts of D3 charge.]{\label{f:threetwo:a}
\includegraphics[scale=0.58]{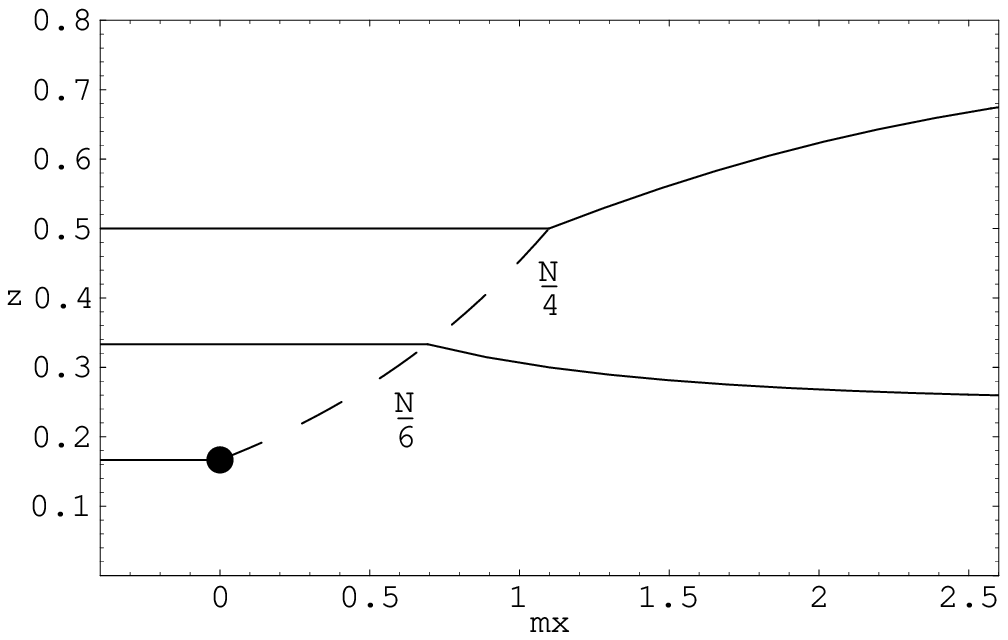}}
\subfigure[The three D5s carry 1/4, 1/4, and 1/2 of the D3 charges; the two
carry 1/3 and 2/3.]{\label{f:threetwo:b}
\includegraphics[scale=0.58]{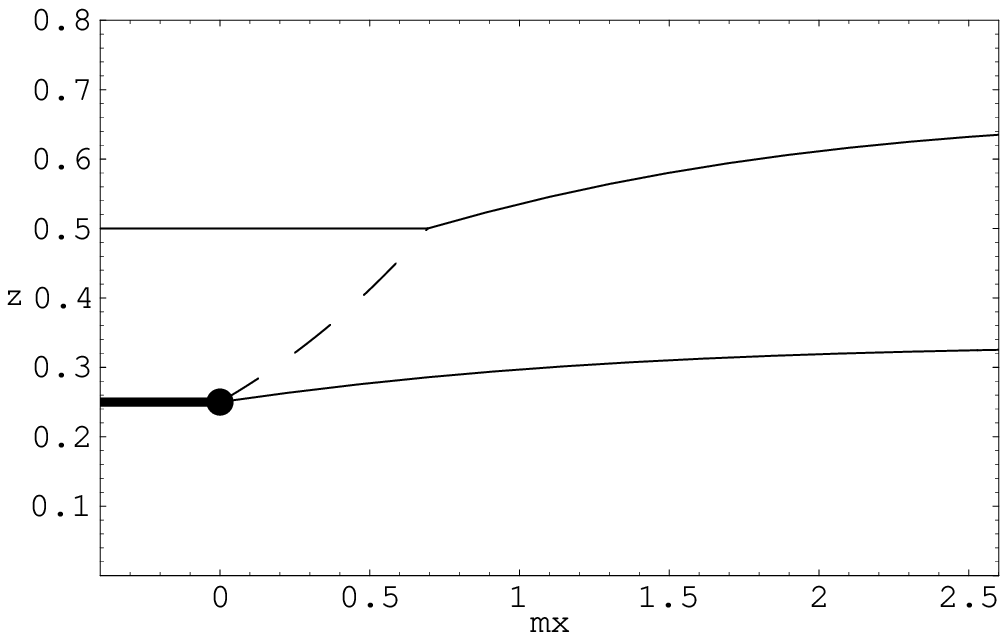}}
\caption{\label{f:threetwo}BPS domain walls for three D5-branes going to two 
D5 branes.  In both cases, the vacuum with three 5-branes has a
smaller magnitude superpotential than the vacuum with two.  In part 
\ref{f:threetwo:b}, the two smaller 5-branes at 
negative $x$ are chosen to carry the same D3 charge.  Thick lines indicate
two D5-branes.}
\end{figure}

A final illustrative case to consider is a domain wall interpolating between
Coulomb vacua with two and three D5-branes respectively.  If none of the
three 5-branes is larger than both of the two 5-branes, then 
a BPS domain wall can be constructed
similar to the first domain wall discussed above (see figure 
\ref{f:threetwo} for some BPS domain walls).  However, in cases where one
of the three D5-branes is larger than both of the D5- branes in the other 
vacuum, there appears to be no BPS domain wall constructed out of real
solutions to the BPS equation.  If, for example, the three D5-brane vacuum
has a smaller magnitude superpotential,
the domain wall would require a zero-5-brane
with negative D3 charge, which we have seen are ruled out.
Similarly, if the three D5 vacuum has a greater magnitude superpotential,
there would be at least one 5-brane junction with more D5-branes at larger $x$,
violating the condition of equation (\ref{cond}).

\begin{figure}[t]
\centering
\includegraphics[scale=0.8]{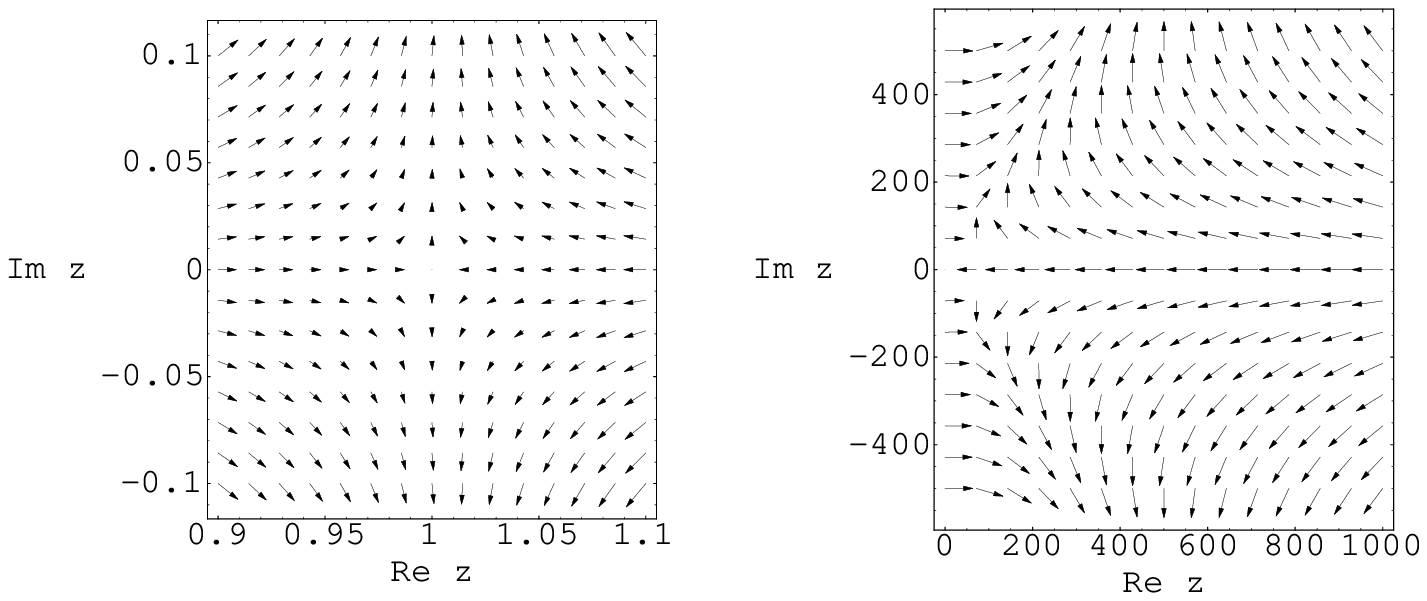}
\caption{\label{f:compfield}A plot of $\partial_{mx} \psi_{1,2}$ at two 
axis scales; axes are given by $z=(\psi_1+ i\psi_2)/\psi_v$.
It does not seem that any solution leaving the horizontal axis from 
$\psi_1/\psi_v = 1$ would ever return to the real axis.}
\end{figure}

Might there be BPS
domain walls between these vacua with complex values of $\psi$?  Considering
the BPS equation (\ref{realbps}) for $\psi = \psi_1+i\psi_2$ (for $\psi_{1,2}$
real), we have
\begin{eqnarray}
\frac{1}{m}\partial_x \psi_1 &=& \psi_1 - \frac{1}{\psi_v}\left(
\psi_1^2-\psi_2^2\right)\label{bps_1}\\
\frac{1}{m}\partial_x\psi_2 &=&\left( 2\frac{\psi_1}{\psi_v} -1\right)\psi_2
\, .\label{bps_2}
\end{eqnarray}
The boundary conditions on the vacuum branes at positive $x$ require that
they do not twist (that $\psi$ remains real).  To see this, consider a
perturbation of equations (\ref{bps_1},\ref{bps_2}) for a single 5-brane 
around $\psi_v$ at $x=+\infty$
and note that such a perturbation vanishes at $\infty$ only if it is real.  
Then equation (\ref{bps_2}) requires the solution to remain real.  For the
vacuum branes at negative $x$, any imaginary perturbation is unstable 
(because $\psi_1 \rightarrow \psi_v$ as $x\rightarrow -\infty$), leading
one to suspect that there is no BPS domain wall between these vacua.  Figure
\ref{f:compfield} shows a vector field of $(1/m) \partial_x (\psi_1\ \psi_2)$
in the $\psi_1,\ \psi_2$ plane, which seems to indicate that an imaginary
perturbation would not flow back to the real axis.  It is possible that some
other type of 5-brane plays a role in these ``complex'' domain walls; however,
such a configuration, if it exists, would be difficult to find.

It is, however, not altogether surprising that some pairs of
vacua do not have BPS domain walls; consider one vacuum with D5-branes 
of D3 charge  $(12,16,19,24)N$ D3 and another vacuum with D5-branes of D3 
charges respectively 
$(9,10,15,27)N$.  These two vacua have the same superpotential,
so any BPS domain wall between them would be tensionless.  On the other hand,
the vacua are not identical, so there must be some positive tension due to
brane bending in any domain wall.  Thus, there is no BPS domain 
wall\footnote{Thanks to I. Bena and M. Patel for discussions of this point.}.  
It is worth noting, though, that higher order effects in the 5-brane charge or
in the $1/N$ expansion might or might not lift the degeneracy of these vacua 
and permit a BPS domain wall.  Also, it is known that $\mathcal{N}=1$ 
supersymmetric $SU(N)$ theories with matter in the fundamental representation
do not have BPS domain walls if the matter mass is too large
\cite{Smilga:1998vs,Smilga:1997cx,Smilga:1997pf}\footnote{We need not worry
that the value of the mass will alter the spectrum of BPS domain walls in 
our case; since we work with a deformation of a conformal theory, $m$ is the
only length scale.}.  

We can now conjecture several conditions necessary for BPS domain walls
to exist between a pair of Coulomb (or Higgs) vacua, based on the discussion
above.  We should stress that these conditions would be proved if we require
$\psi$ to be real, but we cannot absolutely rule out domain walls with
complex $\psi$.  First, the superpotential must take different values in the
two vacua, $W(x\rightarrow -\infty) \neq W(x\rightarrow \infty)$.  Next,
because of the force-balancing condition, the number of 5-branes must not
increase
as $x$ (more physically, the $|W|$) increases.  And further, the size of the
largest 5-brane (meaning its D3 charge) must not decrease as $x$ ($|W|$) 
increases; this follows because the size of a zero-5-brane, which carries D3
charge from one 5-brane junction to another, increases with $x$.  If we 
considered vacua with a larger number of 5-branes, we would find a larger
number of conditions that must be satisfied; from the brane physics, it would
seem that those conditions could be described as the conditions above applied
to subsets of the 5-branes in the two vacuum states.

\subsection{Domain Walls Between Single 5-Branes}\label{s:singlebranes}
In this section, we consider domain walls between vacua with one 5-brane.
As discussed in section \ref{s:general}, we know that the BPS trajectory along
the domain wall follows a straight line in the complex superpotential plane,
even including discontinuities at 5-brane junctions, and 
$\Omega = \Delta W/|\Delta W|$.  This allows us to solve
the full complex BPS equation (\ref{bps}) numerically; we can solve the cubic
superpotential $W$ for the configuration variable $\phi$ and plot the 
trajectory in configuration space as a function of $W$ on the line segment
from $W(x\rightarrow -\infty)$ to $W(x\rightarrow\infty)$.  If we do this
for the two vacuum branes, we can find the configuration $\phi_0$ where they
intersect to find the BPS domain wall.  For single 5-brane vacua, D3 charge is
automatically conserved at the brane junction, and, for the specified value
of $\Omega$, the phase condition (\ref{phase}) is satisfied up to a sign,
which we check numerically.

One issue is that there are three solutions for $\phi$ as a function of $W$
because the superpotential is cubic.  However, only two of these approach
the vacuum configuration $\phi_v$ as the superpotential goes to the vacuum
value $W_v$ (see eqn. (\ref{vacua})).  Those two solutions correspond
to the two solutions $\psi_>$ and $\psi_<$ discussed above, essentially.
We can choose whichever solution will intersect with the other 5-brane.

(An additional check on numerical solutions is provided by the argument of
$\partial_x \phi$ at the vacuum state.  By considering linear perturbations
of the BPS equation (\ref{bps}) around the vacuum, we find 
\begin{eqnarray}
\partial_x \delta\phi_1 &=& -m(\Omega_1 \delta\phi_1 + \Omega_2 \delta\phi_2)
\label{pert1}\\
\partial_x \delta\phi_2 &=& -m(\Omega_2 \delta\phi_1 - \Omega_1 \delta\phi_2)
\label{pert2}
\end{eqnarray}
for $\delta\phi_{1,2}$ and $\Omega_{1,2}$ real and imaginary parts 
respectively.  Only one of the eigenvectors has $\delta\phi\rightarrow 0$
for $x\rightarrow \infty$ (or $-\infty$, as desired), so this gives the phase
of $\partial_x \phi$.  This checks numerically.)

\begin{figure}[p]
\centering
\subfigure[String coupling $g=1$.]{\label{f:dto11:a}
\includegraphics[scale=0.58]{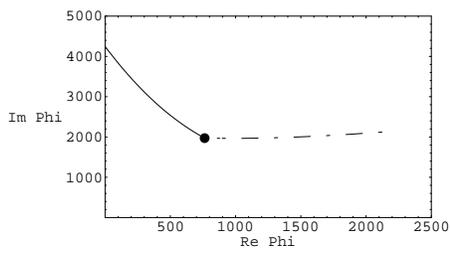}}
\subfigure[String coupling $g=0.1$.]{\label{f:dto11:b}
\includegraphics[scale=0.58]{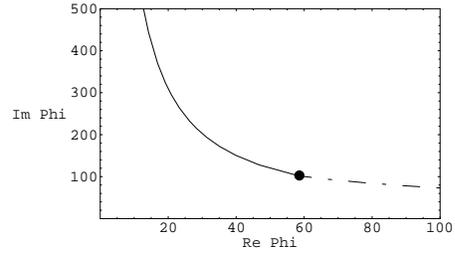}}
\subfigure[String coupling $g=10$.]{\label{f:dto11:c}
\includegraphics[scale=0.58]{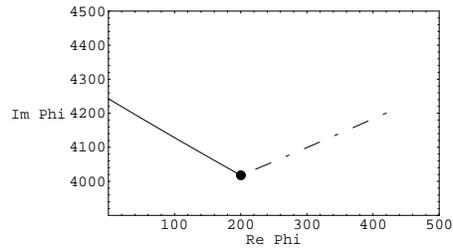}}
\caption{\label{f:dto11} The BPS domain walls interpolating between a D5-brane 
and a $(1,1)$5-brane.  The axes are the real and imaginary $\phi$ axes
(as in figure \ref{f:dns});
the D5 is a solid line, and the $(1,1)$5-brane is the dot-dashed line.  
The dot represents a ball-filling NS5-brane.  Part
\ref{f:dto11:b} is cropped to show the junction more clearly.  The RR scalar
$\theta$ is set to zero.}
\end{figure}

The first case to consider are for a domain wall interpolating between a 
D5-brane and a $(1,1)$5-brane ($(1,k)$5-brane vacua are sometimes called
oblique confining vacua, while the NS5-brane vacuum is the confining vacuum). 
Figure \ref{f:dto11} shows 
graphs of the brane configuration parameter $\phi$ for
a domain wall interpolating between a D5-brane and a $(1,1)$5-brane with
several values of the string coupling $g$.  For small $g$, the $(1,1)$5-brane
looks like an NS5-brane, while for large $g$, it looks like a D5-brane.
Thus, for small $g$, most of the domain wall tension is from brane bending,
while, for large $g$, it is mostly from the 3-ball 5-brane.  The NS5-$(1,1)$ 
domain wall is S-dual to this case, and the other D5/NS5- to $(1,k)$5-brane
domain walls are similar.  All these 
configurations satisfy the force-balancing condition, 
and they are the only BPS domain walls found between each pair
of vacua.

\begin{figure}[t]
\centering
\subfigure[$\theta=(0.01)2\pi$, D5 to NS5]{\label{f:dns:a}
\includegraphics[scale=0.58]{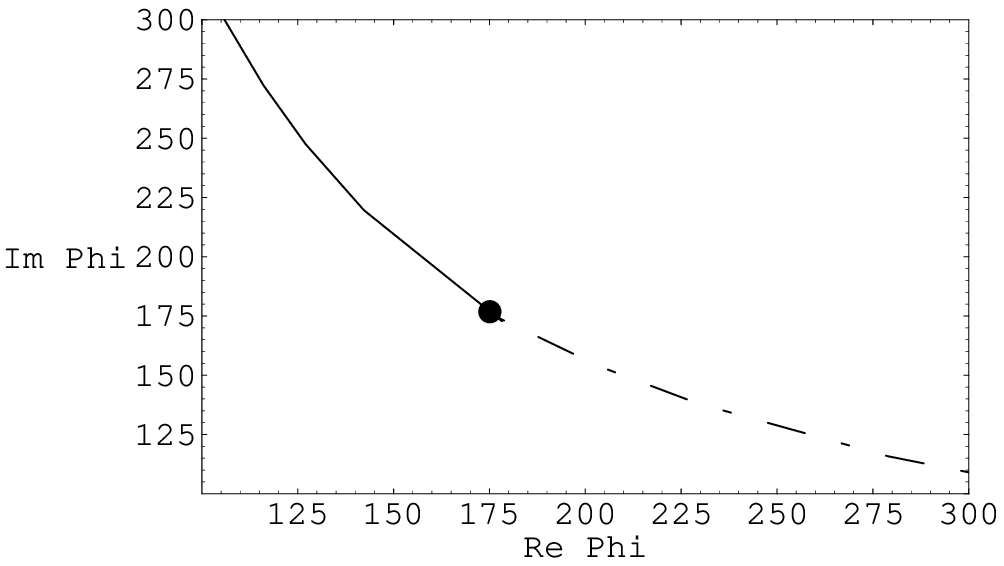}}
\subfigure[$\theta=(-0.01)2\pi$, anti-D5 to NS5]{\label{f:dns:b}
\includegraphics[scale=0.58]{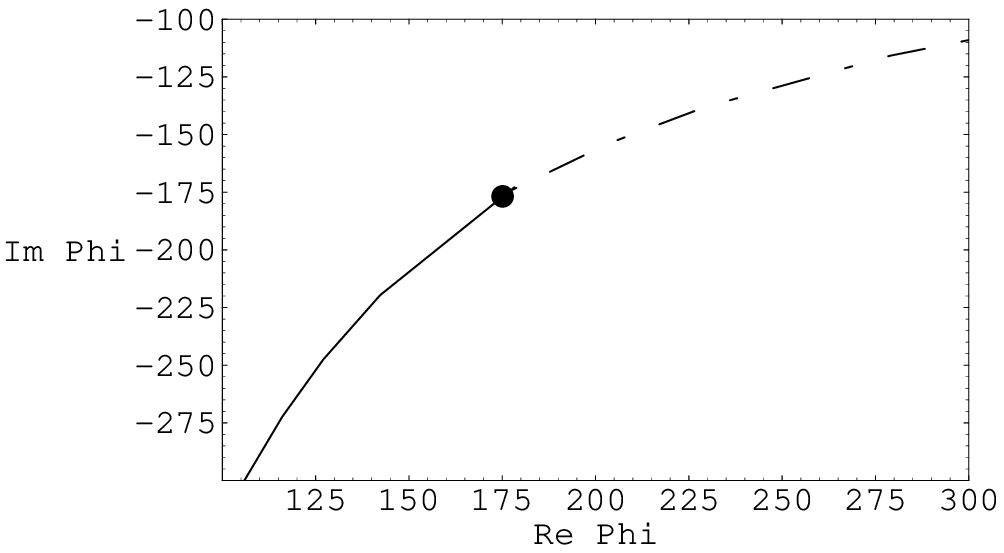}}
\caption{\label{f:dns}BPS domain walls between the Higgs and confining vacua
with $\theta \neq 0$.  The D5-brane is a solid line, the NS5 dot-dashed, and
$g=1$.  Large dots are ball-filling 5-branes.  Note that these plots are the
complex conjugates of each other.}
\end{figure}

\begin{figure}[t]
\centering
\subfigure[D5 brane flow]{\label{f:dnsvect:a}
\includegraphics[scale=0.58]{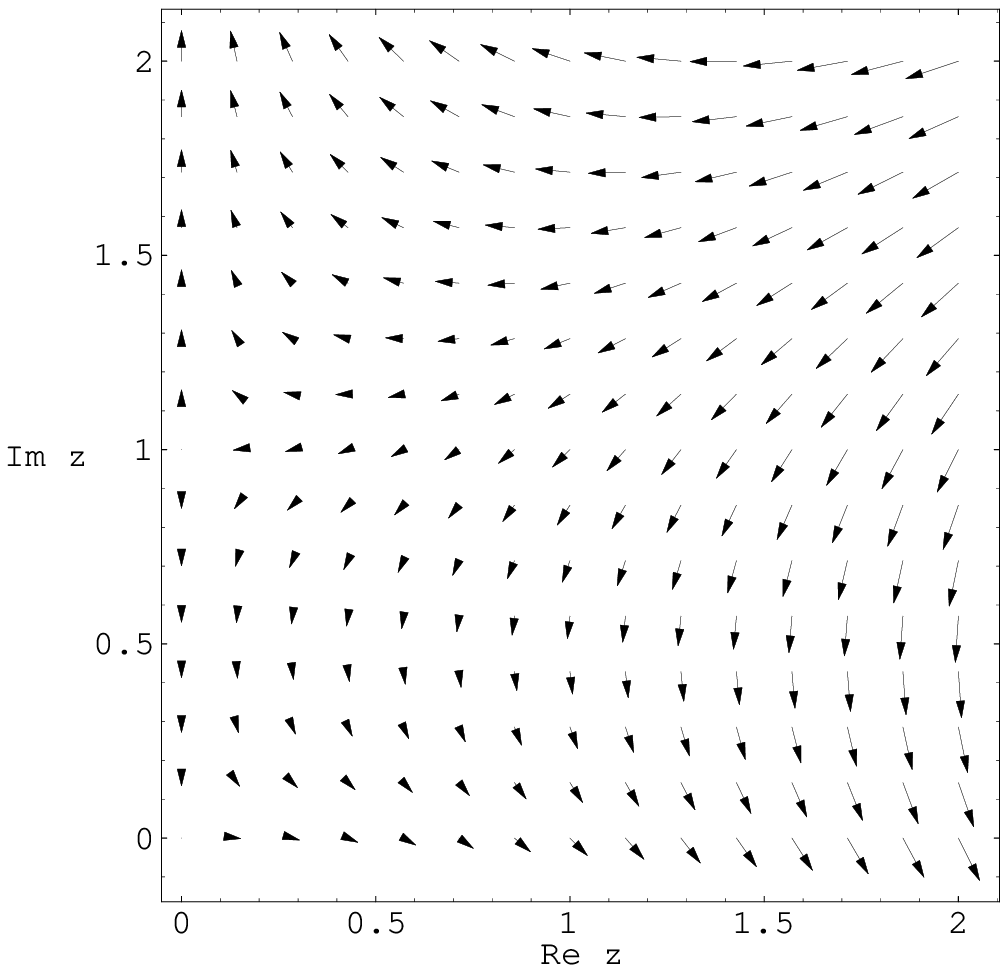}}
\subfigure[NS5 brane flow]{\label{f:dnsvect:b}
\includegraphics[scale=0.58]{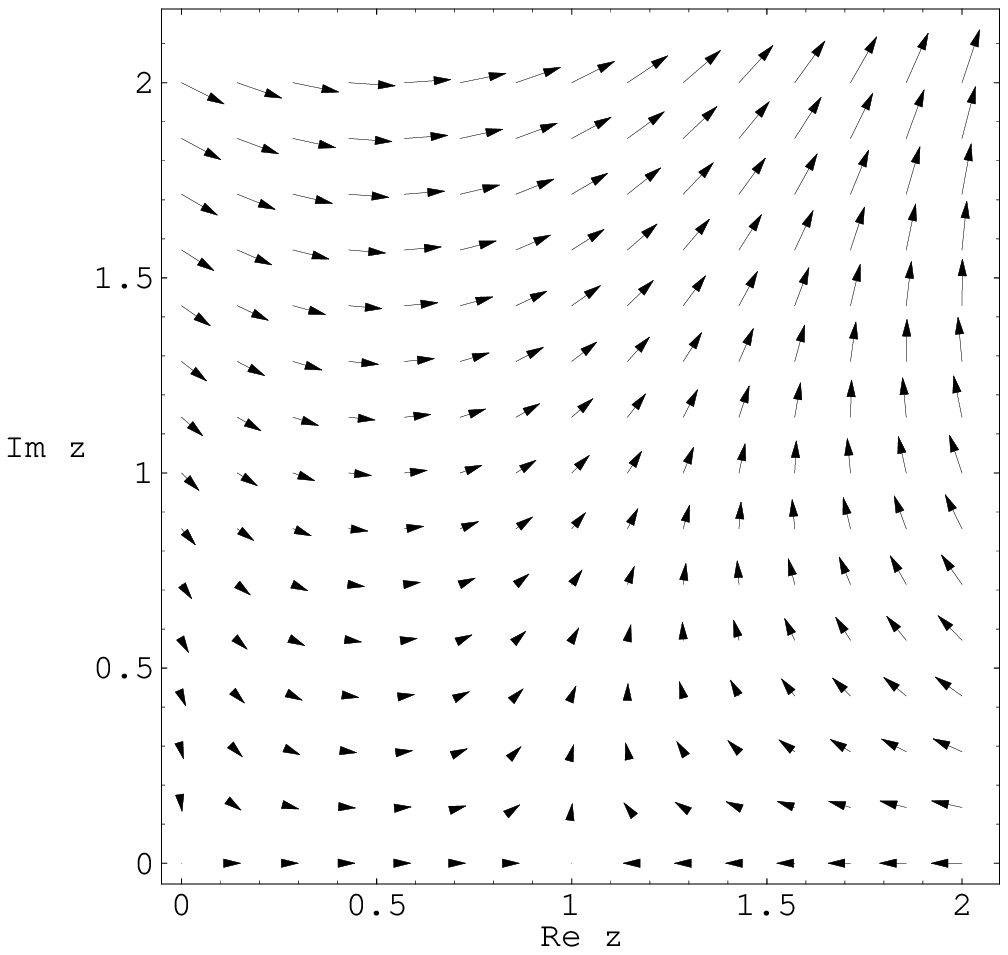}}
\caption{\label{f:dnsvect}Vector flow fields for the D5/NS5-brane bending
with $\Omega=1$, $g=1$, $\theta=0$.  The axes are given by $z=\phi/|\phi_v|$, 
where the vacuum state spheres have the same size.
Note that the D5 flow could conceivably approach the real axis, but, once it
converted to an NS5-brane, it would be swept away from the real axis.}
\end{figure}

The domain wall interpolating between a D5-brane vacuum and an NS5-brane
vacuum is an interesting case.  
For a value of the RR scalar $\theta >0$,
we can find the domain wall just as before (see figure \ref{f:dns:a}).  For
$\theta <0$, on the other hand, the D5 and NS5 branes intersect with opposite
values of $\phi$, meaning that we can interpret the domain wall as 
interpolating between an anti-D5-brane and an NS5-brane (or between a D5- 
and an anti-NS5-brane) (see figure \ref{f:dns:b}).  
For $\theta =0$, though, the domain wall does not seem to
exist.  We can understand this from the BPS equation solutions found in
section \ref{s:onecharge}; 
for the appropriate value of $\Omega=1$ in this case, 
both of the 5-branes follow the solution $\psi_<$ (eqn. (\ref{lesssol})).
(See equations (\ref{pert1},\ref{pert2}) to see that the appropriate
perturbations are along the real and imaginary axes for the NS5- and D5-branes
respectively.)
Thus, the NS5(D5)-brane stays on the real (imaginary) 
$\phi$ axis, and they intersect only at the origin, which their solutions
reach only in an infinite distance.  Figure \ref{f:dnsvect} shows a vector 
field of the BPS equation flow for both the D5- and NS5-brane, and they seem
not to intersect.  Another argument that the D5- and NS5-brane (with 
$\theta =0$) have no BPS domain wall connecting them is that such a 
domain wall would seem to violate the 5-brane/anti-5-brane symmetry of the 
physics.  This second argument should hold even when higher order corrections
are taken into account.  It is also notable that the tension
of this domain wall would scale as $N^4$ in the 't Hooft limit, indicating
that it may not exist \cite{Dorey:2000fc}.

Physically, we seem to have a case of BPS spectrum restructuring in different
regions
of parameter space, as was discussed recently by \cite{Ritz:2000xa}.  In their
language, $\theta =0$ is the curve of marginal stability for the domain wall
between the D5- and NS5-branes.  For positive $\theta$, the ``stable'' BPS 
object is a domain wall between D5- and NS5-branes, while for negative $\theta$
the BPS domain wall is between an anti-D5- and an NS5-brane.  At $\theta=0$,
if the analogy with the results of \cite{Ritz:2000xa} holds, 
the domain wall supposedly becomes a composite of widely separated BPS 
solitons, presumably the domain walls between the D5/NS5-branes and the
``vacuum'' at $\phi=0$.  As mentioned above, most sources consider the 
$\phi=0$ vacuum to be unphysical, and we certainly do not have a good 
description of its physics if it does exist, since corrections due to the
5-brane charge and quantum string physics would play a large role there.
Another possibility is that near shell corrections or quantum effects glue
together the D5- and NS5-branes at a small size, leaving a very thick but
finite size domain wall between the Higgs and confining vacua.  That would be  
a specific
example of the transformation of 5-brane charge at zero size conjectured
in \cite{Polchinski:2000uf}.

\subsection{Multiple 5-branes of Different Types}\label{s:multi}

To find domain walls between the most general pairs of vacua, namely those
with several 5-branes of different types, we will typically need to use the
numerical methods of section \ref{s:singlebranes} above.  These are 
applicable basically without modification, since the superpotential is
additive over different 5-branes and the BPS brane-bending equations are
decoupled for the separate branes.  It is typically difficult to find 
explicit solutions in the most general cases, because three or more curves
in the complex plane will generically not all intersect at a point. Such
general BPS domain walls, if they exist, require extra 5-branes, 
similar to the zero-5-branes in section \ref{s:onecharge} 
but without 
exact solutions to the BPS equation that make it possible to find the
domain walls.  Here, we will discuss two cases with extra symmetry that allow
us to find BPS domain walls.

First, we can discuss domain walls connecting D5-branes to $(1,k)$5-branes.
Assuming that all of the 5-branes in each of the vacua have the same number
of D3 charges (that is, not all of the gauge symmetry is broken), 
these domain walls are more or less rescaled versions
of those discussed above.  These domain walls are uncomplicated because the 
vacua are essentially single 5-branes with multiple 5-brane charges.

\begin{figure}[t]
\centering
\includegraphics[scale=0.58]{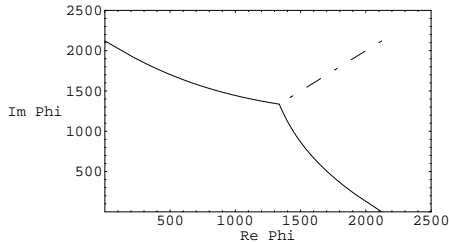}
\caption{\label{f:dns11}The domain wall for a D5- and NS5-brane going to a 
$(1,1)$5-brane with $g=1$ and $\theta=0$.  There is no 5-brane ball. The
D5 and NS5 are solid lines, the $(1,1)$5-brane is dot-dashed.}
\end{figure}

The case of a D5 and NS5 each with $N/2$ D3-brane charges going to a 
$(1,1)$5-brane also has a special symmetry; it is self-S-dual 
for $\theta=0$ and $g=1$, and there
is no ball-filling 5-brane.  This domain wall is shown in figure
\ref{f:dns11} for $g=1$.  For other values of the string coupling, the three
vacuum branes never all have the same $\phi$, so any domain wall would be
more complicated, as in more general cases.  However, one would expect that
BPS domain walls would exist at least for a range of $g$ near $1$ because
one does exist at that special value of the coupling.

\section{Summary}\label{s:summary}
\subsection{Domain Walls}\label{s:sumdw}
In summary, we have discussed BPS domain walls in the string theory dual of
the $\mathcal{N}=1^*$ $SU(N)$ theory.  Using a small-bending approximation
for the vacuum state 5-branes, we found, as in \cite{Polchinski:2000uf}, 
that the 5-branes bend
independently of each other (that is, independently of the configuration
variables of the other 5-branes, including the metric factor $Z$).
We were also able to establish, using conditions for mechanical equilibrium, 
that the BPS bound for the domain wall tension is the same in the string
theory as in the field theory dual, given that the vacuum superpotentials
are the same.

We then discussed domain wall solutions for a number of pairs of vacua.
For domain walls between Coulomb (and Higgs) vacua, we gave analytic 
solutions for BPS brane bending and demonstrated a general construction of
BPS domain walls.  We gave an example of a non-BPS domain wall
and a mechanism through which it can decay classically to a BPS domain wall.
We were further able to find conditions in which BPS domain
walls do not seem to exist.  For a BPS domain wall between Coulomb vacua
to exist:
\begin{itemize}
\item the vacuum superpotentials must differ,
\item the number of 5-branes must not increase as $|W|$ increases, and
\item the size of the largest radius 5-brane must not decrease as 
$|W|$ increases.
\end{itemize}
We also found numerical BPS domain wall solutions interpolating between
the Higgs and oblique confining vacua, as well as for some special cases with
multiple 5-branes.

\subsection{Zero Radius 5-Brane Spheres}\label{s:zero}
Importantly, we also encountered physics involving the zero radius 
configuration of the 5-brane spheres.  In section \ref{s:onecharge}, we 
used zero-5-branes to carry D3 charge without carrying 5-brane charge; the
vacuum state for such a zero-5-brane would have a collapsed sphere, if it
exists.  We re-emphasize that a vacuum state for the zero-5-brane need not
be physical for zero-5-branes to occur in non-vacuum configurations, such as
domain walls.  We also found that the domain wall between the Higgs (D5) and 
confining (NS5) vacua with zero RR scalar seems not to exist in this 
approximation.  One interpretation of this is that the zero-size state (which
is a minimum of the potential) does exist and that $\theta=0$ is a
curve of marginal stability on which the D5/NS5 domain wall decomposes into
two domain walls.  However, recent literature is of the opinion that
the zero-size state is unphysical (as supported by the string exclusion
principle \cite{McGreevy:2000cw}\footnote{Thanks to J. McGreevy for a 
discussion of this point.}).  
In that case, it is possible that effects due to the
5-brane charge or quantum string physics connect the D5- and NS5-branes at a
finite size.  At this time, the physics behind such a transition is not
understood and remains a point of interest for future study.

It may also be interesting to compare the possible zero-size vacuum of brane
physics to the controversial chirally symmetric Kovner-Shifman vacuum of
supersymmetric $SU(N)$ theory \cite{Kovner:1997im}.  For example, both appear
as minima of effective potentials that describe the theory around a particular
vacuum state (for a discussion of the Veneziano-Yankielowicz Lagrangian 
for $\mathcal{N}=1$ Yang-Mills theory \cite{Veneziano:1982ah}
and
an extension of it, see \cite{Gabadadze:1998bi}) which may or may not be valid
near the zero-size or chirally symmetric vacuum respectively.  
Assuming they exist, the zero-size and Kovner-Shifman vacua do have some
at least superficial similarities; first, they both have a zero superpotential,
as opposed to the (oblique) confining vacua.  
They also would both have BPS domain
walls connecting them to the oblique confining vacua in which one real field
varies (in the Kovner-Shifman case, that is the gluino condensate
\cite{Kovner:1997ca}).
It would be 
interesting to calculate the gluino condensate in the zero-size vacuum through
the AdS/CFT correspondence in order to see if it is also chirally symmetric,
but -- if that vacuum exists -- it lies outside of the approximations of
\cite{Polchinski:2000uf}.  However, both of these vacua are generally 
considered to be unphysical (see \cite{Hollowood:1999qn} for a recent critique
of the Kovner-Shifman vacuum).  Perhaps a physical understanding of why one
of these vacua fails to exist (or a determination that it does indeed exist)
would shed light on the physics of the other.

\subsection{Comparison to Field Theoretic Results}\label{s:compare}
It is also possible to connect our results to previously known field theoretic
results.  First, we will compare our results to the recent work of 
\cite{Bachas:2000dx}, which found BPS domain wall configurations interpolating
between Coulomb and Higgs vacua in the $\mathcal{N}=1^*$ theory, 
as in section \ref{s:onecharge} here.  In
general, our results agree (including the solutions to the BPS equation, when
given), but there are two comments to be made here. One
comment is that the configuration (2.14) of \cite{Bachas:2000dx} does not 
connect the Coulomb vacua to the confining vacuum, as we have discussed
extensively above.  Because \cite{Bachas:2000dx} considered only the classical
vacuum structure, they neglected quantum effects that give the confining vacuum
a gluino condensate and non-zero superpotential.  The other comment is
a detailed comparison of our conditions for the existence of BPS domain walls
to the conditions (4.21) (or equivalently (4.19)) of 
\cite{Bachas:2000dx}\footnote{Thanks to 
B. Pioline for pointing out revisions to \cite{Bachas:2000dx} 
regarding these conditions.}.
The condition that the number of 5-branes (that is, the number of $SU(2)$
irreducible representations) not increase with increasing $|W|$ is precisely
equivalent to the condition that $k_1$ not increase (following the notation
of \cite{Bachas:2000dx}).  The condition on the size of the 5-branes is more
difficult to translate.  Suppose that the largest 5-brane in the vacuum with
smaller $|W|$ has D3 charge $p$, which is larger than any of the 5-branes
in the other vacuum.  Then $k_{p-1}(\mbox{larger}\ |W|)=N$, while
$k_{p-1}(\mbox{smaller}\ |W|)=N-q$, where $q$ is the number of 5-branes with
D3 charge $p$, violating the condition that  $k_{p-1}$ not increase. 
Thus, we find agreement with \cite{Bachas:2000dx}, despite the different
physics used to find our results.  We should
note that neither we nor \cite{Bachas:2000dx} have shown that BPS domain walls
with complex values of $\psi$ (corresponding to nonzero gluino condensate)
connecting Coulomb vacua are ruled out, although they seem not to exist
numerically.

Recent studies of BPS domain walls in supersymmetric gluodynamics in the large
$N$ limit have stressed that the BPS tension between adjacent oblique 
confining vacua (such as an NS5- and $(1,1)$5-brane) scales as $N$, while
the natural energy density scales as $N^2$, leading to the conclusion that
the domain wall thickness must vanish as $1/N$ \cite{Kogan:1998dt,
Shifman:1998vf,Dvali:1999pk,Gabadadze:1999pp}.  These studies have also noted
that such a scaling is precisely that expected for a D-brane, in line with
a suggestion by Witten \cite{Witten:1997ep} 
that a domain wall would act as a D-brane
for the QCD string.  As \cite{Polchinski:2000uf} stated,
the domain walls considered here demonstrate that domain walls in
the $\mathcal{N}=1^*$ theory are indeed D-branes.  If we consider, for 
example, the BPS domain wall between an NS5-brane and a $(1,1)$5-brane in
the 't Hooft limit, then the vacuum states differ only at order $1/N$, so the
dominant contribution to the tension is from the ball-filling D5-brane, which
has a vanishing thickness and on which precisely the correct flux tube can
end \cite{Polchinski:2000uf}.

\subsection{Future Directions}\label{s:future}
Finally, we should note a few future directions to take.  In terms of string 
physics, an understanding of the domain wall between D5- and NS5-branes or 
a definitive determination whether it exists would be important in 
understanding
the physics of D-brane spheres at small radius in RR-form backgrounds.  With
the motivation of studying domain walls, 
this work could be extended to BPS domain wall junctions
\cite{Gibbons:1999np,Carroll:1999wr}, which
have been of increasing interest in the literature \cite{Gabadadze:1999pp,
Shifman:1999ri,Gorsky:1999hk,Saffin:1999au,Binosi:1999vb}.  
Another direction might be to eliminate the small-bending
approximation, which would correspond to finding a more general form of
the K\"{a}hler potential for the configuration variable $\phi$ but would make
finding explicit solutions much more difficult.

\section*{Acknowledgements}\label{s:ack}
I am greatly endebted to Joseph Polchinski for many discussions and for
reading this manuscript.
I would also like to thank S. Hellerman, B. Pioline, I. Bena, M. Patel, 
and J. McGreevy for useful discussions.  This material is based upon work 
supported by a National Science Foundation Graduate Research Fellowship.

\bibliographystyle{../utphys}
\bibliography{bdwall}

\end{document}